\documentclass[12pt]{article}
\usepackage{graphicx,epsfig}
\usepackage{color}
\usepackage{amssymb}
\title{Effects of geometrical frustration on ferromagnetism in the Hubbard model  
on the Shastry-Sutherland lattice}
\author{Pavol Farka\v sovsk\'y\\
Institute  of  Experimental  Physics,  Slovak   Academy   of
Sciences\\
Watsonova 47, 040 01 Ko\v {s}ice, Slovakia}
\date{}

\begin{document}
\baselineskip=24pt
\maketitle

\begin{abstract}
The small-cluster exact-diagonalization calculations and the projector 
quantum Monte Carlo method are used to examine the competing effects 
of geometrical frustration and interaction on ferromagnetism 
in the Hubbard model on the Shastry-Sutherland lattice. 
It is shown that the geometrical frustration stabilizes
the ferromagnetic state at high electron concentrations ($n \gtrsim 7/4$),
where strong correlations between ferromagnetism and the shape 
of the noninteracting density of states are observed.
In particular, it is found that ferromagnetism is stabilized 
only for these values of frustration parameters, which lead 
to the single peaked noninterating density of states at the band edge. 
Once, two or more peaks appear in the noninteracting density of states 
at the band egde the ferromagnetic state is suppressed.  
This opens a new route towards the understanding of ferromagnetism
in strongly correlated systems.

\end{abstract}
%\thanks{PACS nrs.:75.10.Lp, 71.27.+a, 71.28.+d, 71.30.+h}

\newpage
\section{Introduction}
Since its introduction in 1963, the Hubbard model~\cite{Hubbard} has 
become, one of the most popular models of correlated electrons 
on a lattice. It has been used in the literature to study a great variety 
of many-body effects in metals, of which ferromagnetism, metal-insulator 
transitions, charge-density waves and superconductivity are the most common
examples. Of all these cooperative phenomena, the problem of ferromagnetism
in the Hubbard model has the longest history. Although the model was
originally introduced to describe the itinerant ferromagnetism 
in narrow-band metals like $Fe, Co, Ni$ and others, it soon turned 
out that the single-band Hubbard model is not the canonical model 
for ferromagnetism. Indeed, the existence of saturated
ferromagnetism has been proven rigorously only for very special limits.
The first well-known example is the Nagaoka limit that corresponds
to the infinite-$U$ Hubbard model with one hole in a half-filled 
band~\cite{Nagaoka}.
Another example, where saturated ferromagnetism has been shown to
exist, is the case of the one-dimensional Hubbard model with
nearest and next-nearest-neighbor hopping at low electron
densities~\cite{M_H}.
Furthermore, several examples of the fully polarized ground state 
have been found on special lattices as are the bipartite lattices 
with sublattices containing a different number of sites~\cite{Lieb},
the fcc-type lattices~\cite{Ulmke,Pandey},
the lattices with long-range electron 
hopping~\cite{UH1,UH2,Fark1,Fark2,Fark3},
the flat bands~\cite{FB2,FB3,FB4,FB5} and the nearly flat-band
systems~\cite{NFB1,NFB2,NFB3,NFB4}.
This indicates that the lattice structure, which dictates the shape
of the density of states (DOS), plays an important role in stabilizing 
the ferromagnetic state.  

Motivated by these results, in the current paper we focus our attention 
on the special type of lattice, the so-called Shastry-Sutherland 
lattice (SSL). The SSL represents one of the simplest systems with 
geometrical frustration, so that putting the electrons on this lattice 
one can examine simultaneously both, the effect of interaction as well as 
the effect of geometrical frustration on ground-state properties of the 
Hubbard model. This lattice was first introduced by Shastry 
and Sutherland~\cite{Shastry} as an interesting example of 
a frustrated quantum spin system with an exact ground state. 
It can be described as a square lattice with 
the nearest-neighbor links $t_1$ and the next-nearest neighbors
links $t_2$ in every second square (see Fig.~1a).
\begin{figure}[h!]
\begin{center}
\includegraphics[width=7cm]{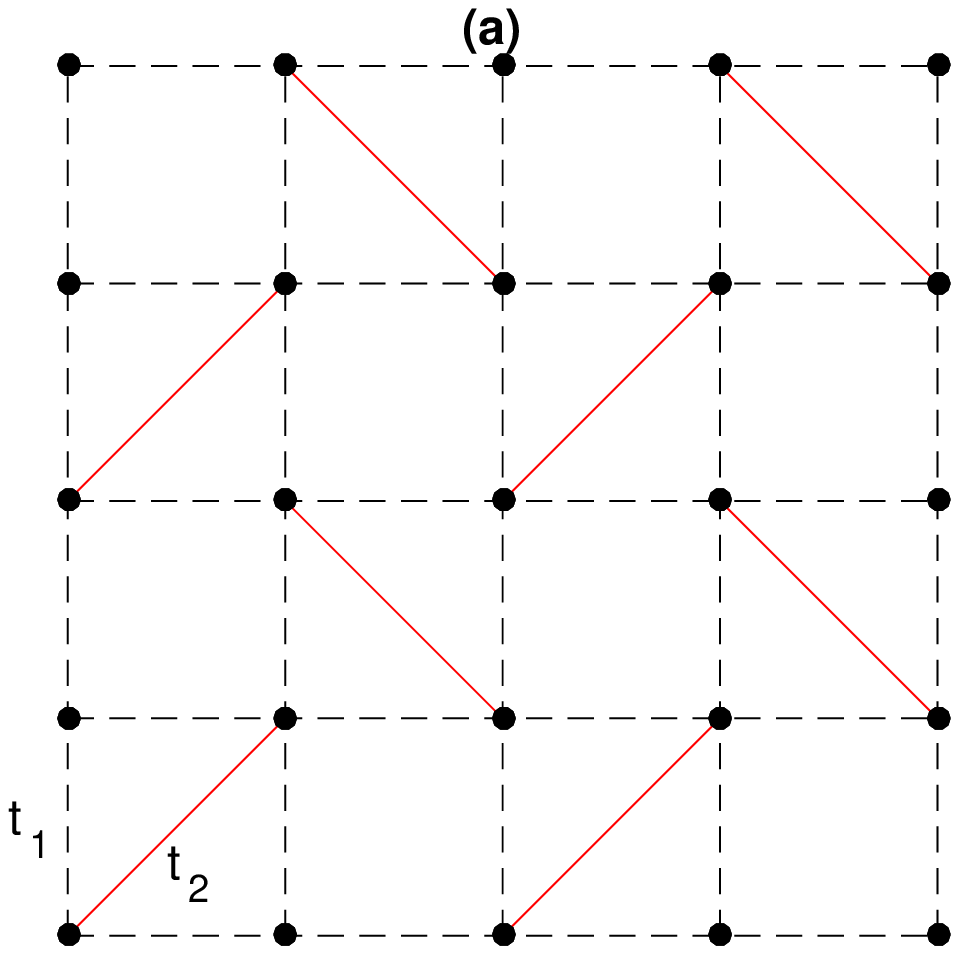}
\includegraphics[width=7cm]{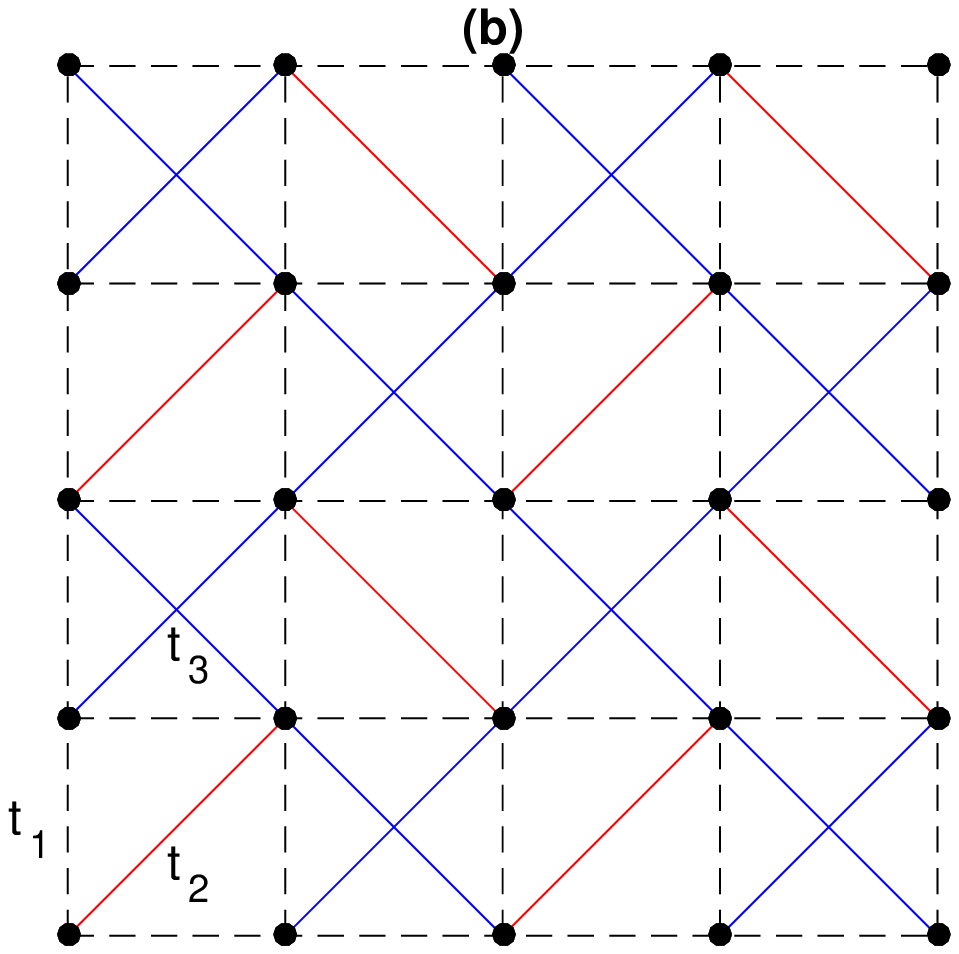}
\end{center}
\caption{\small (a) The original SSL with the first ($t_1$) and second ($t_2$)
nearest-neighbor couplings. (b) The generalized SSL with the first ($t_1$), 
second ($t_2$) and third ($t_3$) nearest-neighbor couplings.   
}
\label{fig1}
\end{figure}
The SSL attracted much attention after its experimental realization
in the $SrCu_2(BO_3)_2$ compound~\cite{Kageyama2}. The observation
of a fascinating sequence of magnetization plateaus (at 
$m/m_s=$1/2, 1/3, 1/4 and 1/8 of  the  saturated  magnetization $m_s$) 
in this material~\cite{Kodama} stimulated further theoretical and experimental 
studies of the SSL. Some time later, many other Shastry-Sutherland  
magnets have been discovered~\cite{Yoshii,Gabani}. In particular,
this concerns an entire group of rare-earth metal tetraborides $RB_4$
($R=La-Lu$). These materials exhibit similar sequences of fractional
magnetization plateaus as observed in the $SrCu_2(BO_3)_2$ compound. 
For example, for $TbB_4$ the magnetization plateau has been found 
at $m/m_s=$2/9, 1/3, 4/9, 1/2  and 7/9~\cite{Yoshii} and for $TmB_4$ 
at $m/m_s=$1/11, 1/9, 1/7 and 1/2~\cite{Gabani}. To describe some of 
the above mentioned plateaus correctly, it was necessearry to generalize 
the Shastry-Sutherland model by including couplings between the third 
and even between the forth nearest neighbors~\cite{pssb}. 
The SSL with the first, second and third nearest-neighbor links is 
shown in Fig.~1b and this is just the lattice that will be used in our 
next numerical calculations.    

Thus our starting Hamiltonian, corresponding to the one band Hubbard
model on the SSL, can be written as
\begin{equation}
H=
-t_1\sum_{\langle ij \rangle_1, \sigma}c^+_{i\sigma}c_{j\sigma}
-t_2\sum_{\langle ij \rangle_2, \sigma}c^+_{i\sigma}c_{j\sigma}
-t_3\sum_{\langle ij \rangle_3, \sigma}c^+_{i\sigma}c_{j\sigma}
+U\sum_{i}n_{i\uparrow}n_{i\downarrow},
\label{Eq3}
\end{equation}
where $c^+_{i\sigma}$ and $c_{i\sigma}$ are the creation and annihilation
operators  for an electron of spin $\sigma$ at site $i$ and
$n_{i\sigma}$ is the corresponding number operator
($N=N_{\uparrow}+N_{\downarrow}=\sum_{i\sigma} n_{i\sigma}$).
The first three terms of (1) are the kinetic energies
corresponding to the quantum-mechanical hopping of electrons
between the first, second and third nearest neighbors and the
last term is the Hubbard on-site repulsion between two electrons
with opposite spins. We set $t_1=1$ as the energy unit and thus
$t_2$ ($t_3$) can be seen as a measure of the frustration strength.   

To identify the nature of the ground state of the Hubbard model 
on the SSL we have used the small-cluster-exact-diagonalization 
(Lanczos) method~\cite{Lanczos} and the projector quantum Monte-Carlo 
method~\cite{PQMC}. 
In both cases the numerical calculations proceed in the folloving steps. 
Firstly, the ground-state energy of the model $E_g(S_z)$ is 
calculated in all different spin sectors $S_z=N_{\uparrow}-N_{\downarrow}$ 
as a function of model parameters $t_2,t_3$ and $U$. Then the resulting
behaviors of $E_g(S_z)$  are used directly to identify the regions
in the parametric space of the model, where the fully polarized state 
has the lowest energy.
%The Lanczos method is used for small clusters (up to L=36 sites)
%and for lager clusters we apply the projector quantum Monte  Carlo method.
%All these calculations are performed using the projector quantum
%Monte-Carlo algorithm~\cite{PQMC}, which applies $exp(-\theta H)$ to a trial 
%wave-function (in our case the solution for $U = 0$). A projector parameter 
%$\theta \sim 30$ suffices to reach well converged values of the observables 
%discussed here. A time slice of $\Delta \theta = 0.05$ was used in general.

\section{Results and discussion}
To reveal possible stability regions of the ferromagnetic
state in the Hubbard model on the SSL, let us first examine
the effects of the geometrical frustration, represented by nonzero
values of $t_2$ and $t_3$, on the behavior of the non-interacting DOS. 
The previous numerical studies of the standard one and two-dimensional 
Hubbard model with next-nearest~\cite{M_H} as well as 
long-range~\cite{Fark1,Fark2,Fark3} hopping showed that just this quantity 
could be used as a good indicator for the emergence of ferromagnetism
in the interacting systems. Indeed, in both models the strong correlation
between ferromagnetism and the anomalies in the noninteracting DOS
are observed. In the first model the ferromagentic state is found 
at low electron concentrations and the noninteracting DOS is strongly
enhanced at the low-energy band edge, while in the second one the ferromagnetic
phase is stabilized at the high electron concentrations and the spectral
weight is enhanced at the high-energy band edge. This leads to the scenario
according to which the large spectral weight in the noninteracting DOS
that appears at the low (high) energy band edges allows for a small
kinetic-energy loss for a state with total spin $S \ne 0$ in reference to one 
with $S=0$. At some finite value of interaction $U$, the Coulomb repulsion
paid for the low-spin states overcomes this energy loss and the high-spin
state becomes energetically favored. The key point in this picture
is the assumption that the shape of the DOS is only weakly modified
as the interaction $U$ is switched on, at least within its low (high)
energy sector.
  
The noninteracting DOS of the $U=0$ Hubbard model on the SSL 
of size $L=200 \times 200$, obtained by exact diagonalization 
of $H$ (for $U=0$) is shown in Fig.~2. The left panels correspond 
to the situation when $t_2 >0$ and $t_3=0$, while the right panels correspond 
to the situation when both $t_2$ and $t_3$ are finite.
\begin{figure}[h!]
\begin{center}
\includegraphics[width=10cm]{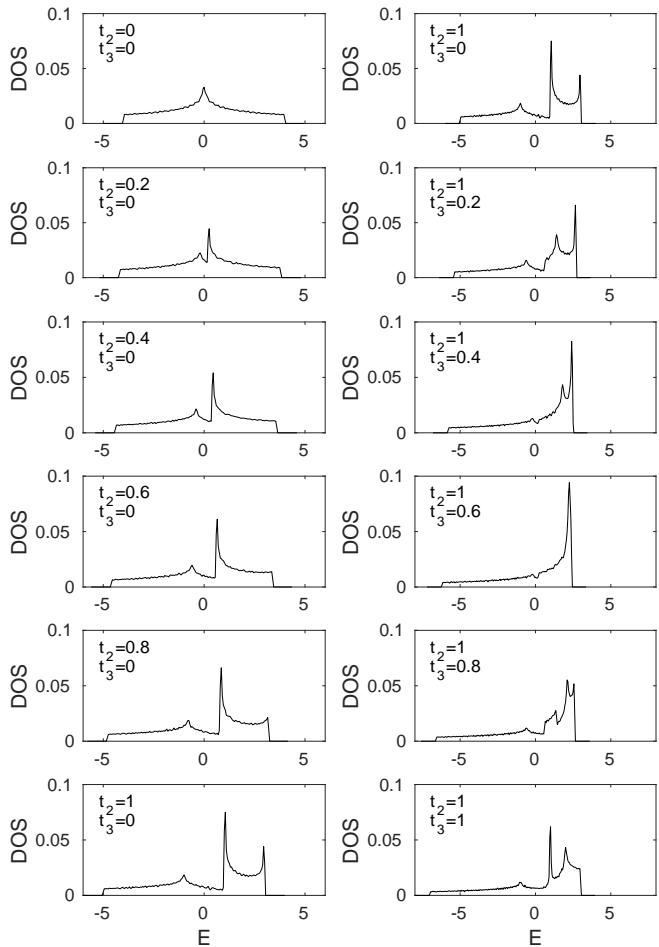}
\end{center}
\caption{\small Non-interacting DOS calculated numerically for different 
values of $t_2$ and $t_3$ on the finite cluster of $L=200 \times 200$ sites.}
\label{fig2}
\end{figure}
One can see that once the frustration parameter $t_2$ is nonzero,
the spectral weight starts to shift to the upper band edge and
the noninteracting DOS becomes strongly asymmetric. Thus taking
into account the above mentioned scenario, there is a real chance
that the interacting system could be ferromagnetic in the limit
of high electron concentrations. To verify this conjecture we have 
performed exhaustive numerical studies of the model Hamiltonian (1)
for a wide range of the model parameters $U,t_2$ and $n$ at $t_3=0$. 
Typical results of our PQMC calculations obtained on finite cluster of 
$L=6\times6$ sites, in two different concentration limits 
($n\leq 1$ and $n>1$) are shown in Fig.~3. 
\begin{figure}[h!]
\begin{center}
\includegraphics[width=7cm]{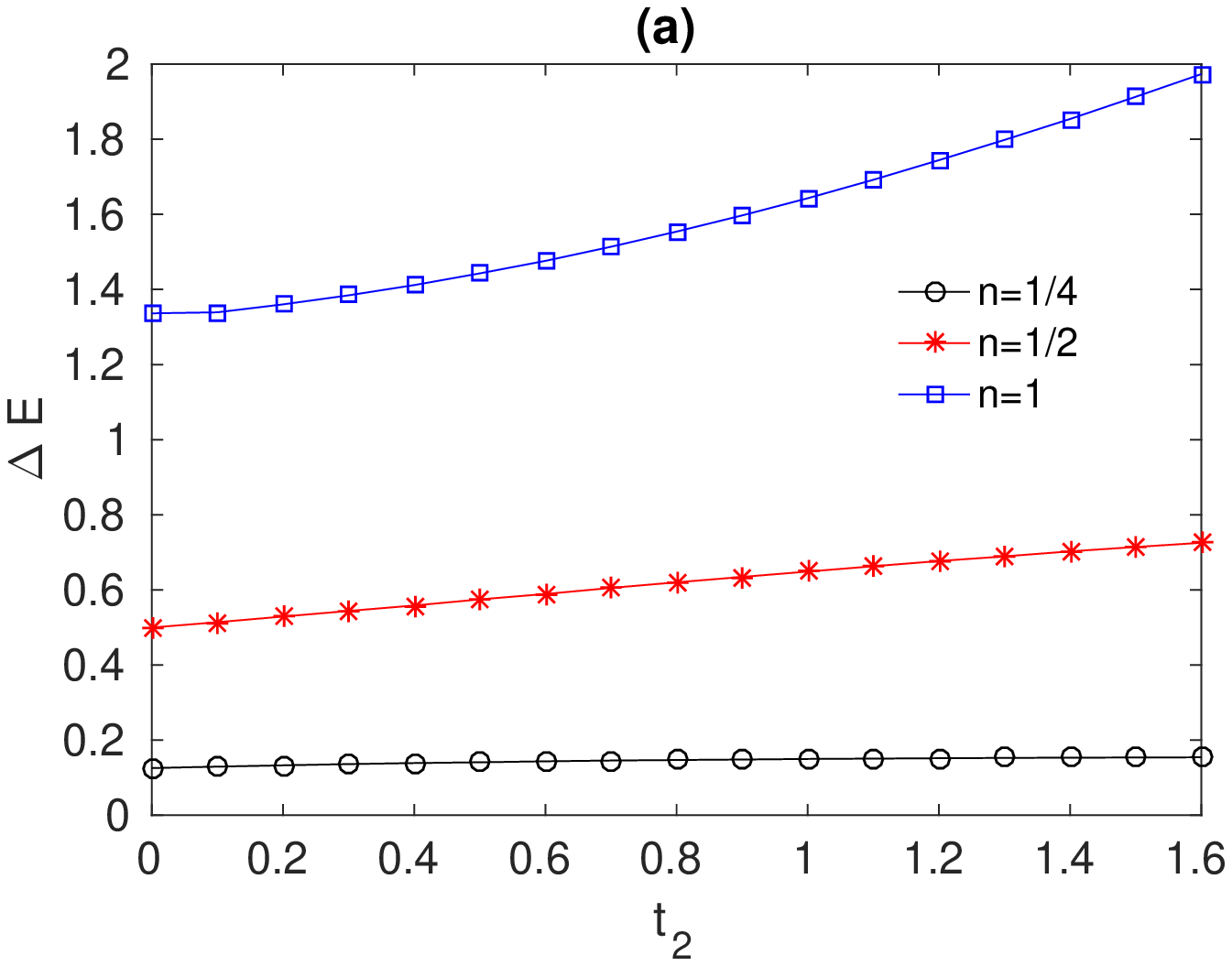}
\includegraphics[width=7cm]{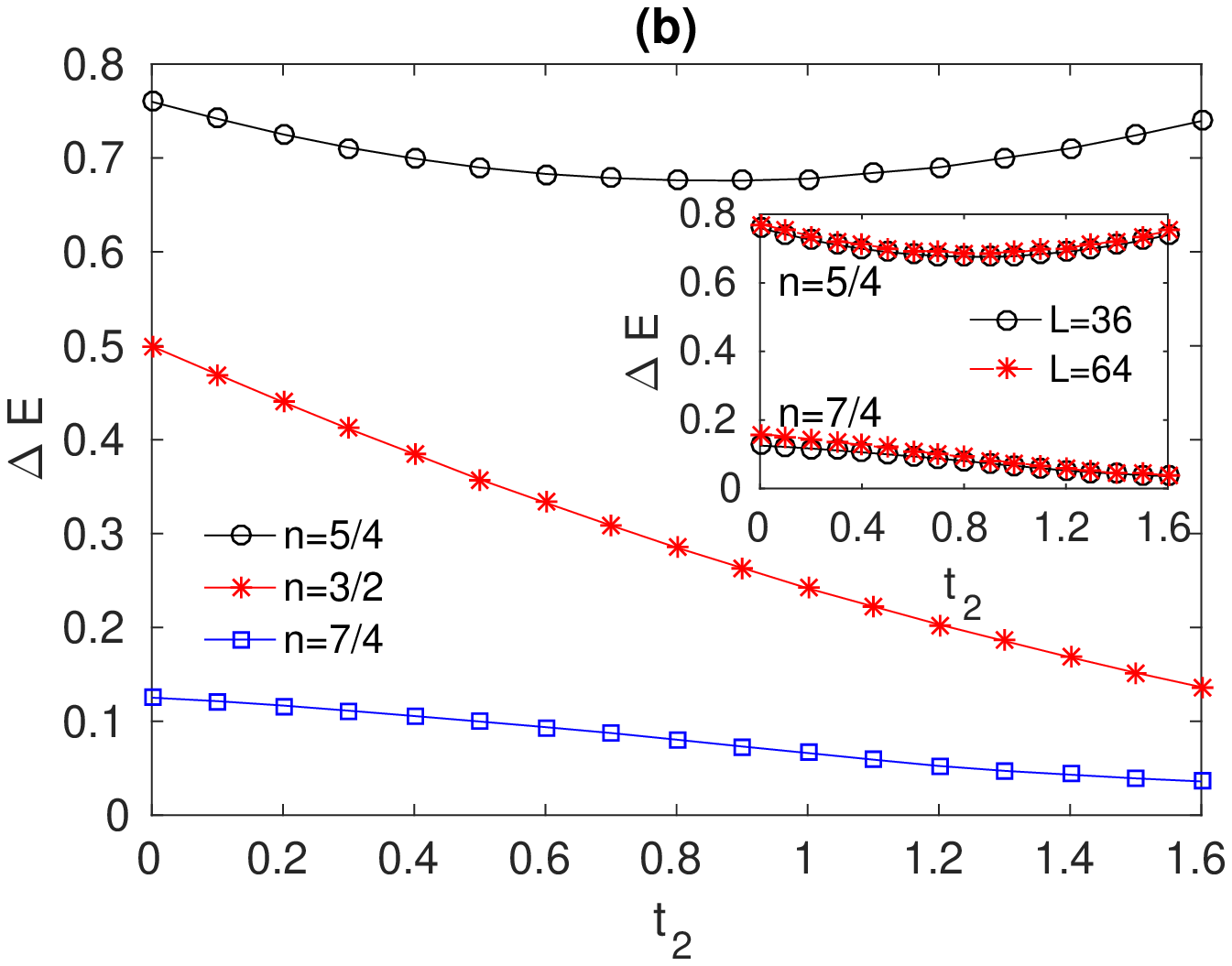}
\end{center}
\caption{\small The difference $\Delta E=E_f-E_{min}$ between the 
ferromagnetic state $E_f$ and the lowest ground-state energy 
from $E_g(S_z)$ as a function of the frustration parameter $t_2$
calculated for $n\leq 1$ (a) and $n>1$ (b) on the finite cluster 
of $L=6 \times 6$ sites ($U=1, t_3=0$). The inset shows $\Delta E$, 
calculated for two different electron densities on clusters of 
$L=6 \times 6$ and $L=8 \times 8$ sites.}
\label{fig3}
\end{figure}
There is plotted the difference
$\Delta E=E_f-E_{min}$ between the ferromagnetic state $E_f$, which
can be calculated exactly and the lowest ground-state energy 
from $E_g(S_z)$ as a function of the frustration parameter $t_2$. 
According to this definition 
the ferromagnetic state corresponds to $\Delta E=0$. It is seen that
for electron concentrations below the half filled band case $n=1$,
$\Delta E$ is the increasing function of $t_2$, and thus
there is no sign  of stabilization of the ferromagnetic state
for $n\leq 1$, in accordance with the above mentioned 
scenario. The situation looks more promising in the opposite 
limit $n>1$. In this case, $\Delta E$ is considerably reduced
with increasing $t_2$, however, this reduction is still insufficient
to reach the ferromagnetic state $\Delta E=0$ for physically 
reasonable values of $t_2$ ($t_2<1.6$) that correspond to the situation
in the real materials. To exclude the finite-size effect, we have also 
performed the same calculations on the larger cluster
of $L=8 \times 8$ sites, but again no signs of stabilization the
ferromagnetic state have been observed (see inset to Fig.~4b).

For this reason we have turned our attention to
the case $t_2>0$ and $t_3>0$. The noninteracting DOS corresponding
to this case is displayed in Fig.~1 (the right panels). These
panels clearly demonstrate that with the increasing value of the 
frustration parameter $t_3$, still a more spectral weight is shifted 
to the upper band edge. A special situation arises at $t_3=0.6$,
when the spectral weight is strongly peaked at the upper band edge.
In this case the nonintercting DOS is practically identical with 
one corresponding to noninteracting electrons with long-range
hopping~\cite{Fark1}.
Since the long-range hopping supports ferromagnetism in the standard
Hubbard model for electron concentrations above the half-filled band 
case~\cite{Fark1,Fark2,Fark3}, we expect that this could be true also 
for the Hubbard model on the SSL, at least for some values of frustration 
parameters $t_2$ and $t_3$. Therefore, we have decided to perform numerical 
studies of the model for a wide range of $t_3$ values at fixed $t_2, U$ 
and $n$ ($t_2=1, U=1, n=7/4$). To minimize the finite-size effects,
the  numerical calculations have been done on two different finite clusters 
of $L=6\times 6$ and $L=8\times 8$ sites. The results of our calculations
for $\Delta E$ as a function of $t_3$ are displayed in Fig.~4a. 
\begin{figure}[h!]
\begin{center}
\includegraphics[width=7cm]{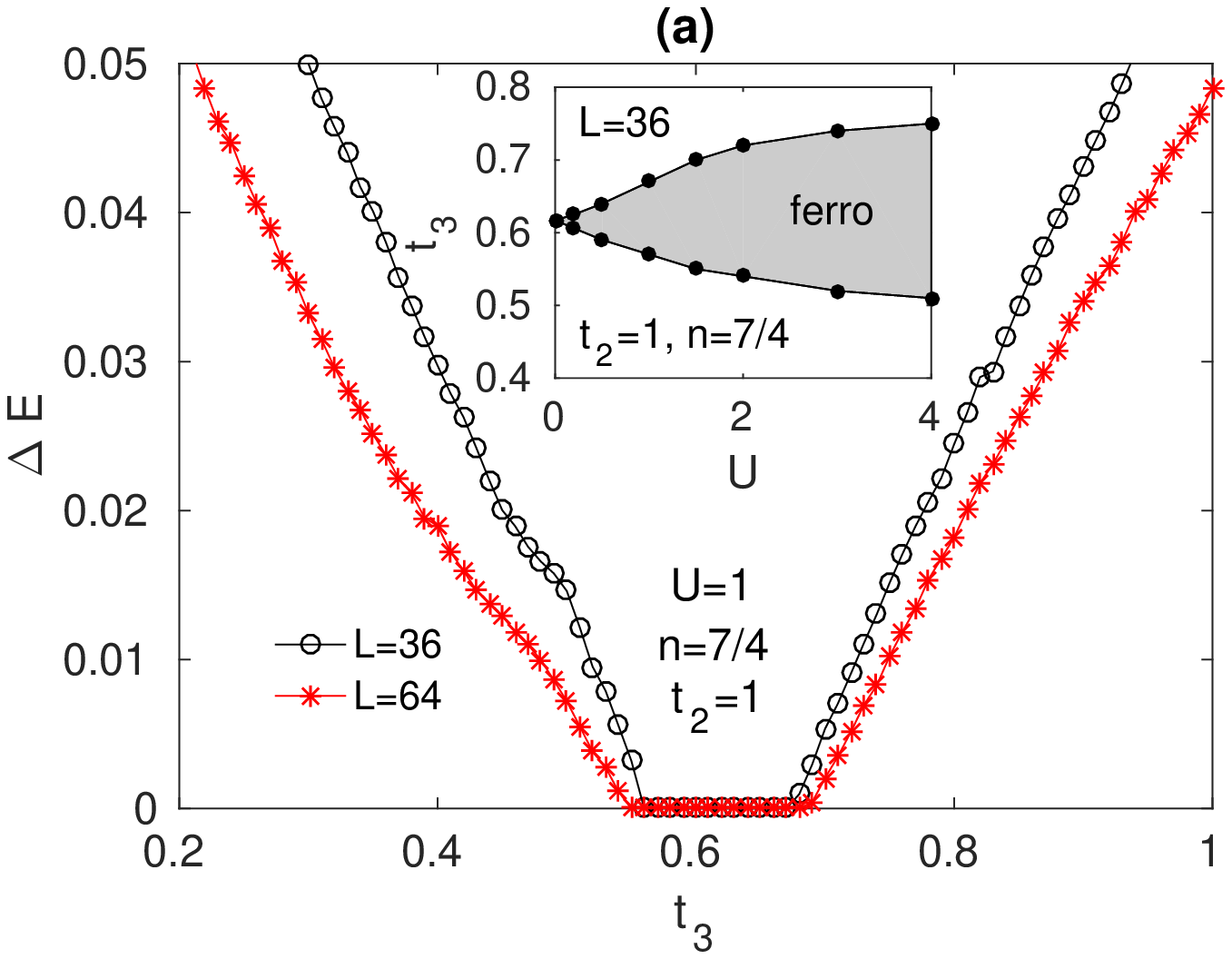}
\includegraphics[width=7cm]{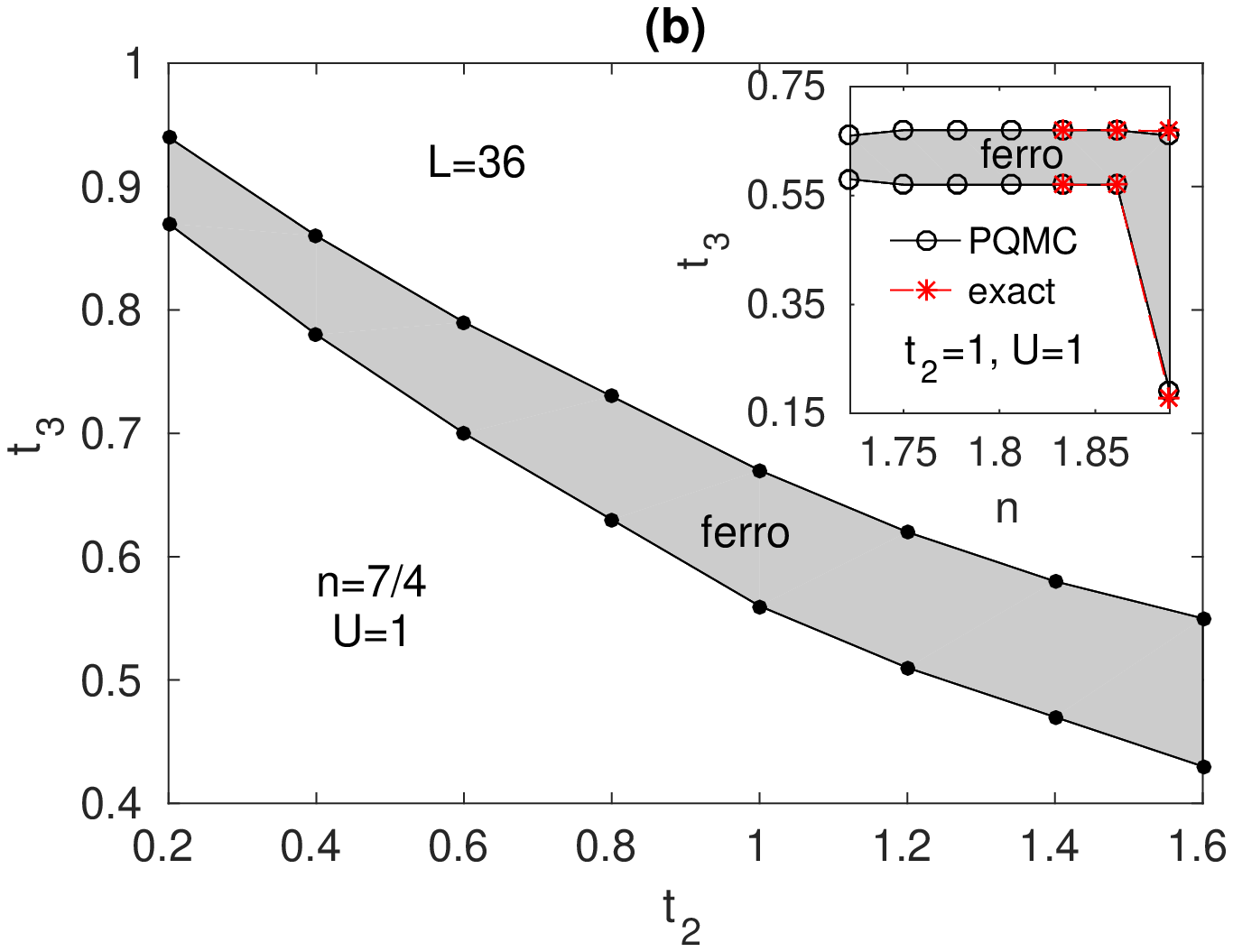}
\end{center}
\caption{\small (a) The difference $\Delta E=E_f-E_{min}$ as a function of 
the frustration parameter $t_3$ calculated for $U=1, t_2=1, n=7/4$ and two 
different finite clusters of $L=6 \times 6$ and $L=8 \times 8$ sites. The
inset shows the ground-state diagram of the model in the $t_3$-$U$ plane.
(b) The comprehensive phase diagrams of the model in the $t_3$-$t_2$
and $t_3$-$n$ plane.}
\label{fig4}
\end{figure}
In accordance with the above mentioned assumptions we find a relatively 
wide region of $t_3$ values around $t_3=0.6$, where the ferromagnetic 
state is stable. It is seen that the finite-size effects on the stability
region of the ferromagnetic phase are negligible and thus these
results can be satisfactorily extrapolated to the thermodynamic limit
$L=\to \infty$. Moreover, the same calculations performed 
for different values of the Hubbard interaction $U$ showed that correlation 
effects  (nonzero $U$) further stabilize the ferromagnetic state and lead 
to the emergence of macroscopic ferromagnetic domain in the $t_3$-$U$ 
phase diagram (see inset to Fig.~4a). This confirms the crucial role
of the Hubbard interaction $U$ in the mechanism of stabilization 
of ferromagnetism on the geometrically frustrated lattice.
In Fig.~4b we have also plotted the comprehensive
phase diagrams of the model in the $t_3$-$n$ as well as $t_3$-$t_2$ plane, 
which clearly demonstrate that the ferromagnetic state is robust 
with respect to doping ($n \gtrsim 7/4$) and frustration. 

To check the convergence of PQMC results we have performed the same 
calculations by the Lanczos exact diagonalization method. Of course, 
on such a large cluster, consisting of $L=6 \times 6$ sites, we were 
able to examine (due to high memory requirements) only several electron 
fillings near the fully occupied band ($N=2L$). The exact diagonalization 
and PQMC results for the width of the ferromagnetic phase 
obtained on finite cluster of $L=6 \times 6$ sites, for three
different electron fillings from the high concentration limit 
($N=66,67,68$), are displayed in the inset to Fig.~4b and 
they show a nice convergence of PQMC results.    

Let us finally turn our attention to the question of possible connection
between ferromagnetism and the noninteracting DOS that has been discussed
at the beginning of the paper. Figs.~4a and 4b  show, that for each finite $U$
and $n$ sufficiently large ($n \gtrsim 7/4$), there exists a finite interval 
of $t_3$ values, around $t_3\sim 0.6$, where the ferromagnetic state is 
the ground state of the model. To examine a possible connection 
between ferromagnetism and the noninteracting DOS, we have 
calculated numerically the noniteracting DOS for several different 
values of $t_3$ from this interval and its vicinity. The results
obtained for $U=1, n=7/4$ and  $t_2=1$ are displayed in Fig.~5.      
\begin{figure}[h!]
\begin{center}
\includegraphics[width=10cm]{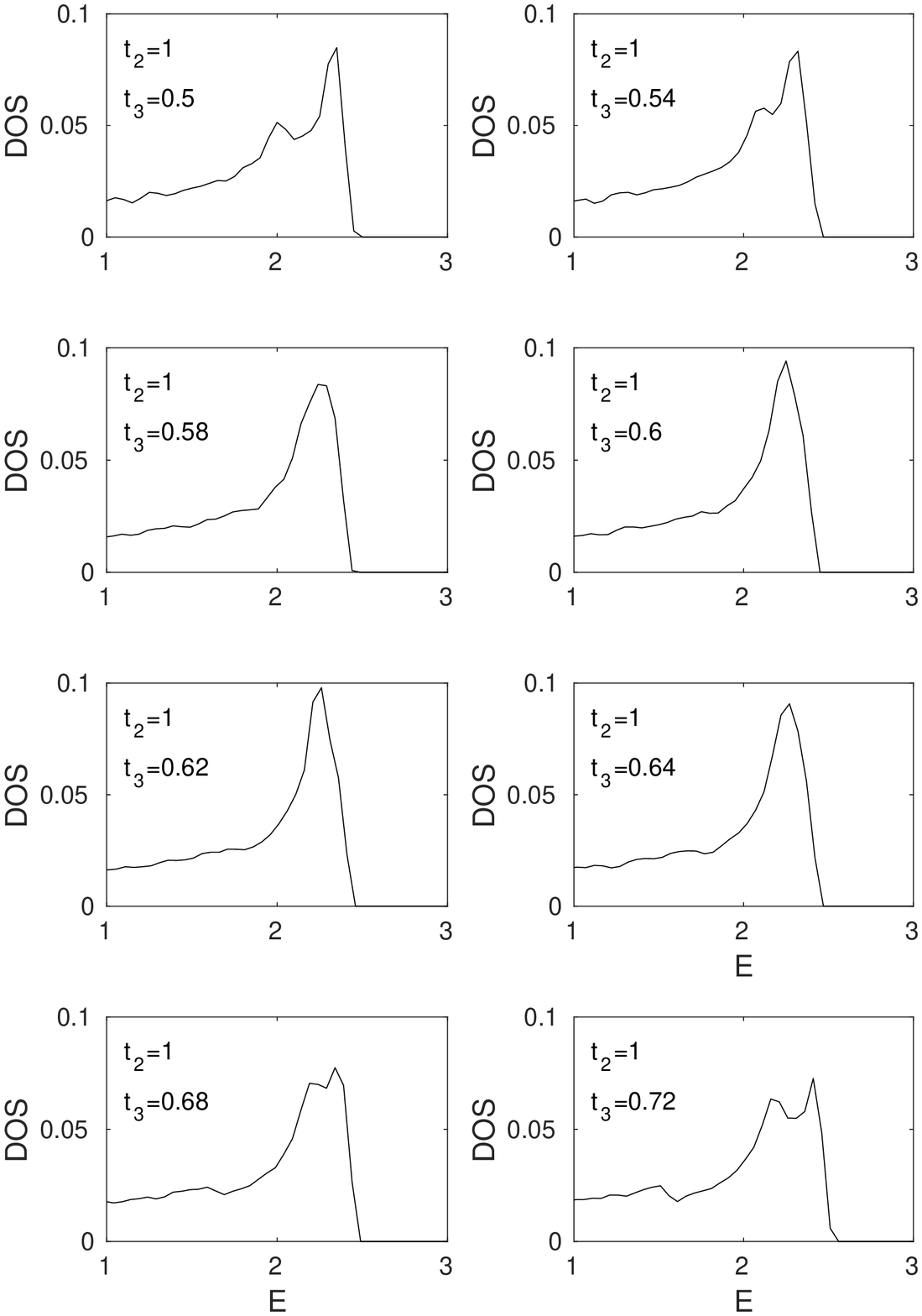}
\end{center}
\caption{\small Non-interacting DOS calculated numericallly for $t_2=1$ 
and different values of $t_3$ (near $t_3=0.6$) on the finite cluster 
of $L=200 \times 200$ sites.}
\label{fig5}
\end{figure}
Comparing these results with the ones presented in Fig.~4a for 
the stability region of the ferromagnetic phase at the same values
of $U,n$ and $t_3$, one can see that there is an obvious correlation 
between the shape of the noninteracting DOS and ferromagnetism. 
Indeed, the ferromagnetic state is stabilized only for these values 
of frustration parameters $t_2,t_3$, which lead to the single peaked 
noninterating DOS at the band edge. Once, two or more peaks appear 
in the noninteracting DOS at the band egde (by changing $t_2$ or $t_3$), 
ferromagnetism is suppressed.

In summary, the small-cluster exact-diagonalization calculations 
and the PQMC method were used to examine possible mechanisms leading 
to the stabilization of ferromagnetism in strongly correlated systems 
with geometrical frustration. Modelling such systems by the Hubbard model 
on the SSL, we have found that the combined effects of geometrical 
frustration and interaction strongly support the formation of the 
ferromagnetic phase at high electron densities. The effects of geometrical 
frustration transform to the mechanism of stabilization of ferromagnetism 
via the behaviour of the noninteracting DOS, the shape of which is determined 
uniquely by the values of frustration parameters $t_2$ and $t_3$. We have 
found that it is just the shape of the noninteracting DOS near the band edge 
(the single peaked DOS) that plays the central role in the stabilization 
of the ferromagnetic  state. Since the same signs have been observed also 
in some other works (e.g., the Hubbard model with nearest and next-nearest 
neighbor hopping, or the Hubbard model with long range hopping), it seems 
that such a behaviour of the noninteracting DOS near the band edge should 
be used like the universal indicator for the emergence of ferromagnetism 
in the interacting systems.

\vspace{0.5cm}
This work was supported by the Slovak Research and Development Agency (APVV)
under Grant APVV-0097-12 and  ERDF EU Grant under the contract No.
ITMS26210120002 and ITMS26220120005.

\end{document}